\documentclass[12pt,preprint]{aastex}
\pdfoutput=1
\usepackage{emulateapj5}
\usepackage{natbib} 
\usepackage{onecolfloat}
\usepackage{apjfonts}
\usepackage{subfigure}

\def\aap{A\&A} 
\def\apj{ApJ} 
 
\def\apjl{ApJL} 
\def\mnras{MNRAS}

\def\gca{Geochimica et Cosmochimica Acta}

\newcommand{\be}{\begin{equation}} 
\newcommand{\ee}{\end{equation}} 
\newcommand{\ba}{\begin{eqnarray}} 
\newcommand{\ea}{\end{eqnarray}} 
\newcommand{\nn}{\nonumber \\}

\shorttitle{Scaling relation in two situations of extreme mergers}
\shortauthors{Rasia et al.}
\begin{document}   
\twocolumn[%

\title{Scaling relation in two situations of extreme mergers}  
\author{
E. Rasia\altaffilmark{1,2}, 
P. Mazzotta\altaffilmark{3,4},
A. Evrard\altaffilmark{1},
M.Markevitch\altaffilmark{3}, 
K. Dolag\altaffilmark{5},    
M. Meneghetti\altaffilmark{6},  
}

\altaffiltext{1}{ Department of Astronomy, University of Michigan, 500 Church St., Ann Arbor, MI 48109-1120, USA, rasia@umich.edu}
\altaffiltext{2}{ Fellow of Michigan Society of Fellows }
\altaffiltext{3}{ Harvard-Smithsonian Centre for Astrophysics, 60 Garden Street, Cambridge, MA 02138, USA}
\altaffiltext{4}{ Dipartimento di Fisica, Universit\`a di Roma Tor Vergata, 
via della Ricerca Scientifica 1,\\ I-00133, Roma, Italy }
\altaffiltext{5}{ Max-Planck-Institut f\"ur Astrophysik, Karl-Schwarzschild Strasse 1, D-85748,\\ Garching bei M\"unchen, Germany}
\altaffiltext{6}{ INAF, Osservatorio Astronomico di Bologna, via Ranzani 1, I-40127, Bologna, Italy }

\begin{abstract}  
Clusters of galaxies are known to be dynamically active systems, yet X-ray studies of the low redshift population exhibit tight scaling laws.   In this work, we extend previous studies of this apparent paradox using numerical simulations of two extreme merger cases, one is a high Mach number (above 2.5) satellite merger similar to the ``bullet cluster'' 
and the other a merger of nearly equal mass progenitors.  
Creating X-ray images densely sampled in time, we construct $T_X$, $M_{gas}$, and $Y_X$ measures within $R_{500}$ and compare to the calibrations of Kravtsov et al. (2006).  We find that these extreme merger cases respect the scaling relations, for both intrinsic measures and for measures derived from appropriately masked, synthetic Chandra X-ray images.  
The masking procedure plays a critical role in the X-ray temperature calculation while it is irrelevant in the X-ray gas mass derivation.
Mis-centering up to 100 kpc does not influence the result. 
 The observationally determined radius $R_{500}$  might conduce to systematic shifts in $M_{gas}$, and $Y_X$ 
 which increase the total mass scatter. 
\end{abstract}

\keywords{cosmology: miscellaneous -- methods: numerical -- galaxies: clusters: general -- X-ray: hydrodynamics -- scaling relations.}  

]

\section{INTRODUCTION}   \label{sec:intro}

The comparison between the observed mass function over cosmic time
with the theoretical expectation constrains various cosmological parameters such as the total mass density, the dark energy equation of state and the normalization of the power spectrum  \citep{vikh.etal.09, mantz.etal.09}.   Such a comparison requires a model for relating observed cluster quantities to the underlying masses of the halo that support them.

The works cited above use the intracluster medium (ICM) of clusters as a proxy for mass.   Estimates of the total cluster mass can be derived from a hydrostatic assumption.  Recent computational work indicates that such estimates are likely to be biased  10-20\% low, on average, with a slightly larger dispersion, relative to the true values \citep{rtm,rasia.etal.06,nagai.etal.07,piffa&valda,jeltema.etal.08,lau.etal.09}.

The dynamical youth of clusters implies incomplete thermalization of the gas, and some pressure support remains in the form of ICM bulk motions and turbulence.  This result was seen in the first three-dimensional, gas dynamical simulations of cluster formation \citep{evrard90}, and it appears to be independent of the numerical algorithm (Lagrangian or Eulerian) and of the physics so far encoded in simulations.   On the last point, some caution is in order because non-thermal effects, such as cosmic-ray pressure and magnetic fields, that may affect the degree of turbulence \citep{lagana.etal.09} have not yet been extensively studied.

For cosmological studies, the slope, zero-point and scatter in the mass--observable relation can be introduced as additional degrees of freedom that are solved for in the likelihood analysis \citep{majumdar&mohr,lima&hu,cunha&evrard}.  Prior constraints on these parameters, from observations or from simulations, can strongly enhance the statistical power of survey analysis.

Scaling relations of the cluster population are thus an important tool for cluster cosmology.  Recent computational modeling emphasizes the regularity of clusters, with the virial theorem a basic organizing mechanism.  The calibration of dark matter (DM) velocity dispersion, $\sigma_{\rm DM}$, by \citet{evrard.etal.08} indicates that $\sigma_{\rm DM}$ at fixed mass and epoch is close to log-normally distributed with a scatter of $4.3 \pm 0.2 \%$, a finding confirmed by the subsequent study of \citet{lau.etal.09}.   
From the observational side, limits to using velocity dispersion as a mass proxy are imposed by projection effects, which substantially enlarge the scatter between mass and line-of-sight galaxy velocity dispersion \citep{biviano.etal.06}, even when the latter is assumed to trace $\sigma_{\rm DM}$.
Samples of roughly 4000 halos covering a range of redshifts from the Millenium Gas Simulations \citep{stanek.etal.09,short.etal.10} support the log-normal assumption for X-ray observable signals, such as ICM mass $M_{gas}$, temperature $T_X$, and their product $Y_X$ \citep{kravtsov.etal.06}.   The effective scatter in mass for these measures, while model-dependent, is $15\%$ or less for the case of a preheated ICM   \citep{stanek.etal.09}.

While such large samples naturally include systems that span a range of dynamical states, from dynamically old, relaxed objects to dynamically young, merging systems, other studies have explicitly attempted to explore the role of mergers on scaling relations.  
\citet{ricker&sarazin} demonstrate the potential for large excursions in X-ray luminosity and temperature in the case of binary mergers.  
This result contrasts with subsequent work by \citet{ritchie&thomas}, who analyze different mergers scenarios including various impact parameters and two different mass ratios.  These authors show that, while mergers do boost the X-ray luminosity and temperature  \citep[{\sl c.f.}][]{torri.etal.04}, 
they do not substantially increase the $L_X-T$ scatter.  \cite{rowley.etal.04} confirm this result, finding no clear relationship between deviation about the mean $L_X-T$ relation and merger activity or formation time.  They report some evidence for a dependence on the degree of substructure.  \cite{ohara.etal.06}, using both simulations and observations, find no influence on  scaling relations from morphological indicators of dynamical state.  Finally, simulation analysis by \cite{hartley.etal.08} indicates that formation epoch is correlated with location in the $L_X-T$ plane.  They see weak negative curvature in the relation at the bright end caused by a prevalence of recently-formed systems.

On the observational side, analysis of luminosity and entropy scaling relations in the REXCESS sample show that morphologically disturbed systems are offset relative to the whole sample, toward lower luminosities and higher entropies at fixed temperature \citep{pratt.etal.09}.  This appears to be due to a deficit of gas interior to $R_{500}$ \citep{pratt.etal.10}.  

While the scaling between X-ray luminosity and temperature has a long history of investigation \citep[e.g.][]{markevitch98,arnaud&evrard,osmond&ponman,pratt.etal.09}, the large scatter in the (non-core-excised) relation has led to the consideration of other mass proxies, especially the gas thermal energy, $Y_X$, the product of the gas mass,  $M_{gas}$, and temperature, $T_{X}$, derived from X-ray observations.

\citet{kravtsov.etal.06} present the $Y_X$ parameter as mass proxy because of its low scatter and nearly self-similar behavior. Using a small set of idealized binary merger simulations, 
\cite{poole.etal.07} 
confirm the tight scatter of $Y_X$ as a mass proxy, and demonstrate that it is relatively insensitive to merger events and projection. With halos drawn from a large volume simulation, \cite{yang.etal.09} find that the scatter of mass--temperature relation depends on the halo concentration: at fixed mass, highly concentrated halos tend to be colder.  That study did not find any bias between merging and non-merging systems.

The simulation evidence cited above indicates that dynamical state plays less of a role than formation epoch in determining the scaling relation location of a particular cluster.  Still, there remains concern that merging systems may bias cluster cosmology studies based on counts as a function of X-ray luminosity function or temperature \citep{randall.etal.02} or Sunyaev-Zel'dovich decrement  \citep{wik.etal.08}.   Here, we revisit the issue of the effect of mergers, emphasizing both theoretical and observational perspectives and using improved simulations with better resolution and more complex physics than used previously.

In this paper, we test the robustness of scaling relations in two specific merging systems.  From a sample of 25 simulated halos \citep{dolag.etal.09}, we select two as examples of extreme cases which can occur in the universe.  The first is a merger happening at $z=0.4$ between a small object and a massive halo, $M_{500} \sim 2 \times 10^{15} M_{\odot}$.  While the secondary structure has a mass of only one-tenth this mass, its close passage through the core plasma causes a detachment of its baryonic and non-baryonic components.  The Mach number of this powerful encounter is 2.5.  Recent observations indicate that these rare circumstances are possible; witness the famous bullet cluster, 1E0657-56, \citep{markevitch.etal.02,clowe.etal.06} and the massive merger of MACS J0025.4-1222 \citep{bradac.etal.08} .
We refer to this simulated system as the `bullet-like cluster' (BL).
The second situation is a merger happening at a more recent redshift ($z \sim 0.09$) between two halos of similar mass, $M_{500} \sim 1.5 \times 10^{14} M_ {\odot}$.  We refer to it as the `1:1 merger' case (1:1).

With respect to previous studies, improvements in this work include the following:  
(i) our simulated systems evolve in a cosmological environment, therefore;  (ii) our clusters experience multiple interactions with the surrounding structures; (iii) the merging phases are followed with high temporal detail, and; (iv) we employ a rigorous X-ray spectral and imaging analysis instead of approximate formulae to calculate X-ray temperature and luminosity.

The paper is structured as follow. Section 2 offers a description of the simulated clusters.  Section 3 describes the evolution of  intrinsic cluster properties for both mergers.  Results based on X-ray analysis are presented in Section 4, and sources of error are dissected in Section 5.  Section 6 offers our conclusions.

\section{Simulated clusters: sample} 
\label{sec:simul}

\begin{figure}
\begin{center} 
\includegraphics[width=0.2\textwidth]{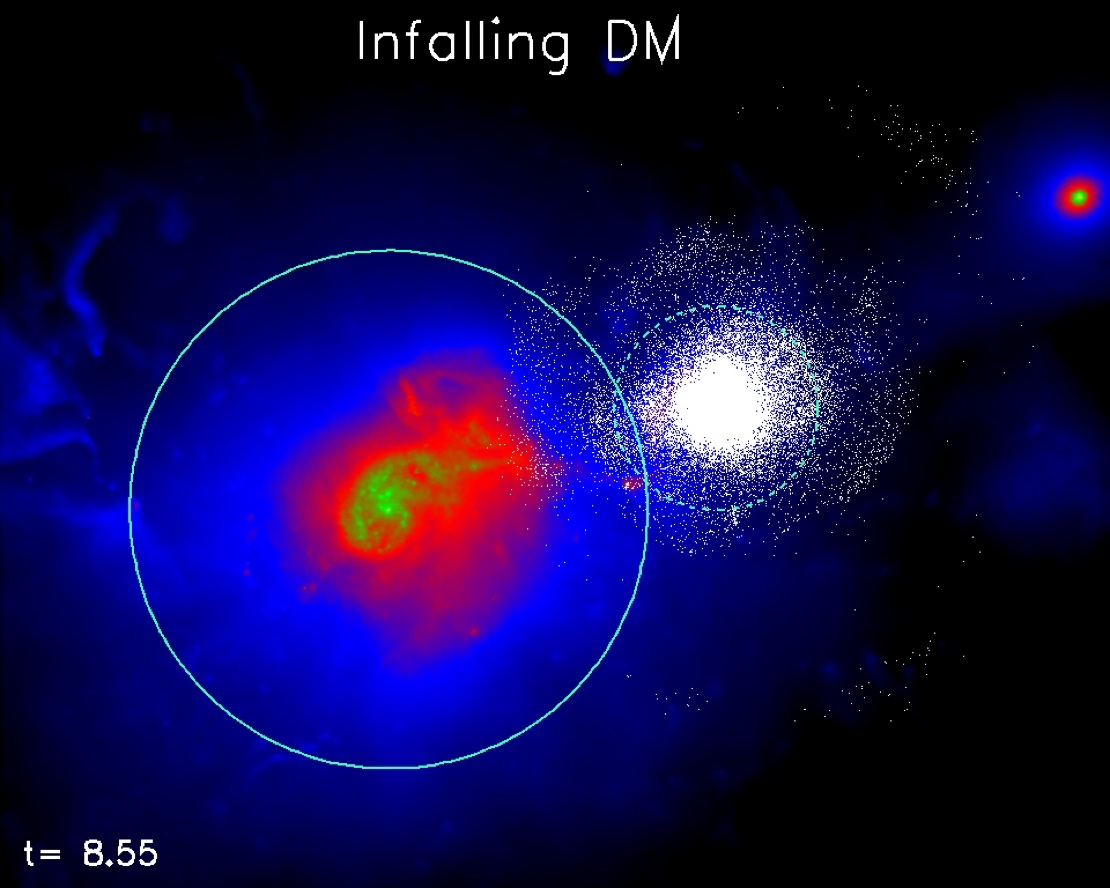}
\includegraphics[width=0.2\textwidth]{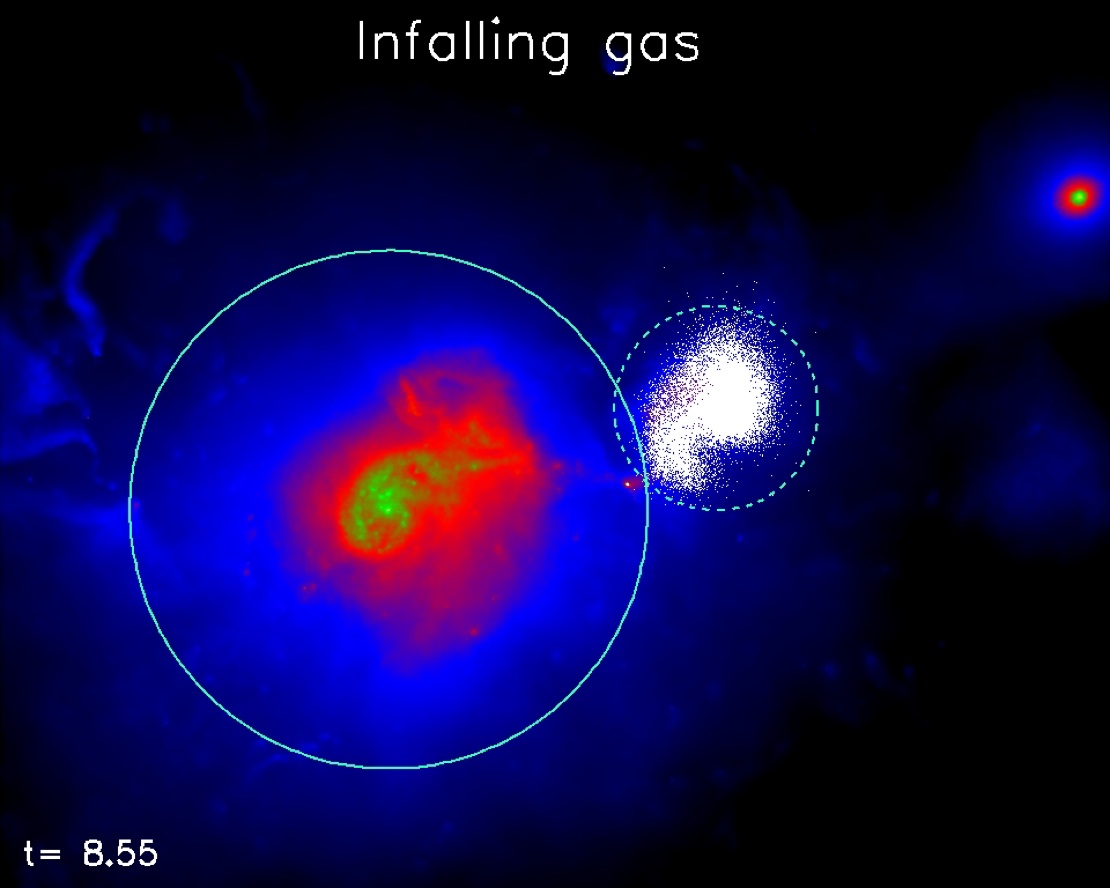}

\includegraphics[width=0.2\textwidth]{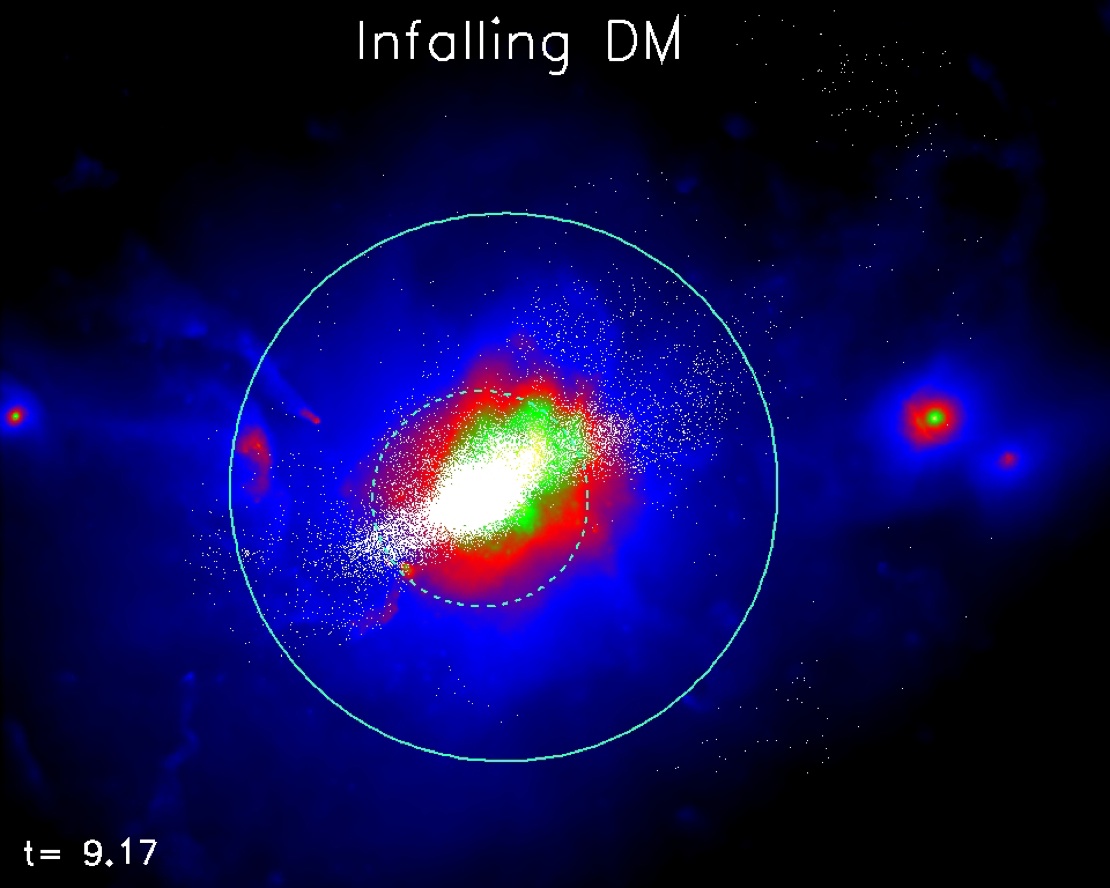}
\includegraphics[width=0.2\textwidth]{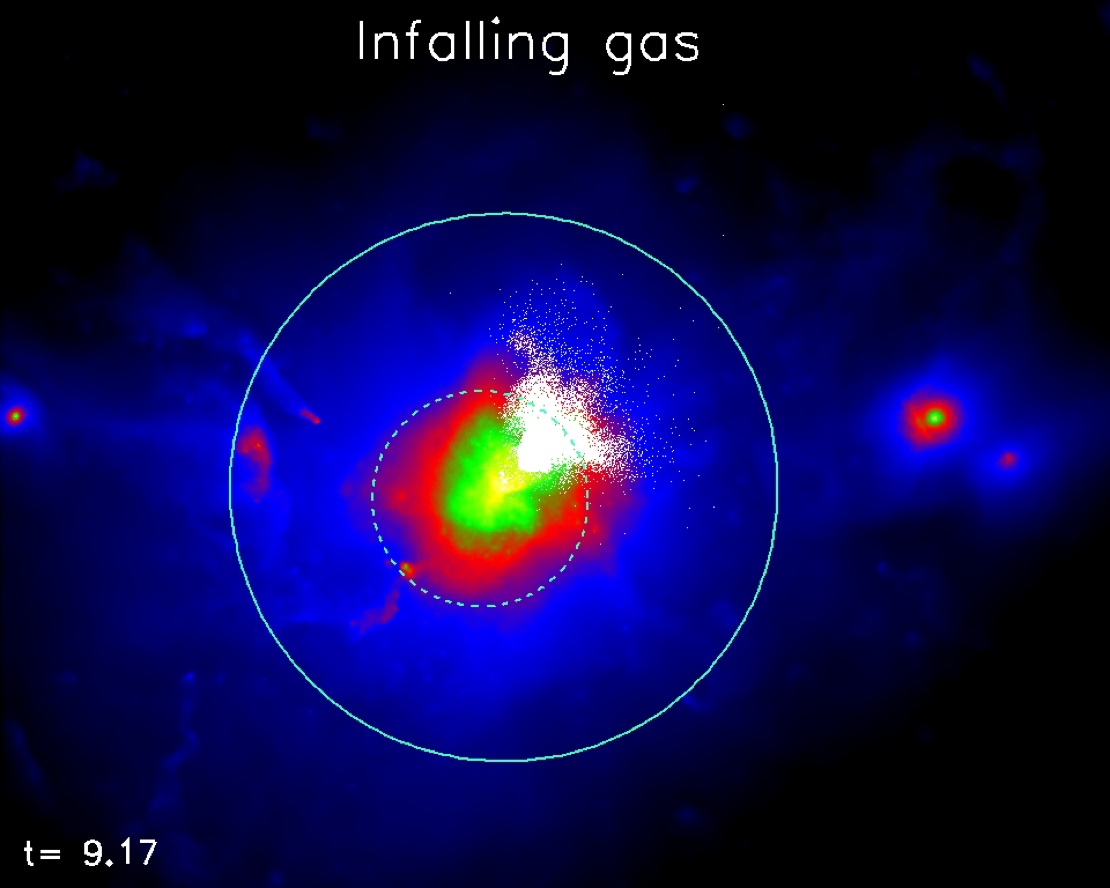}

\includegraphics[width=0.2\textwidth]{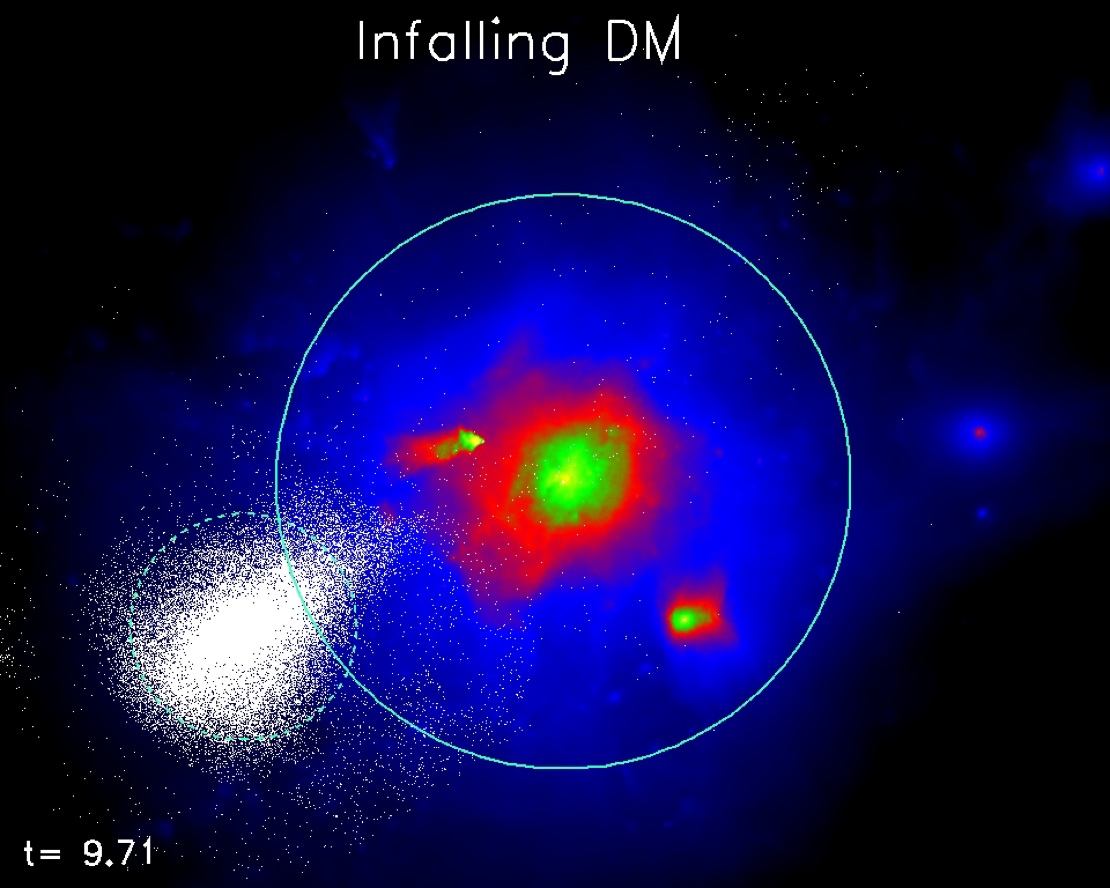}
\includegraphics[width=0.2\textwidth]{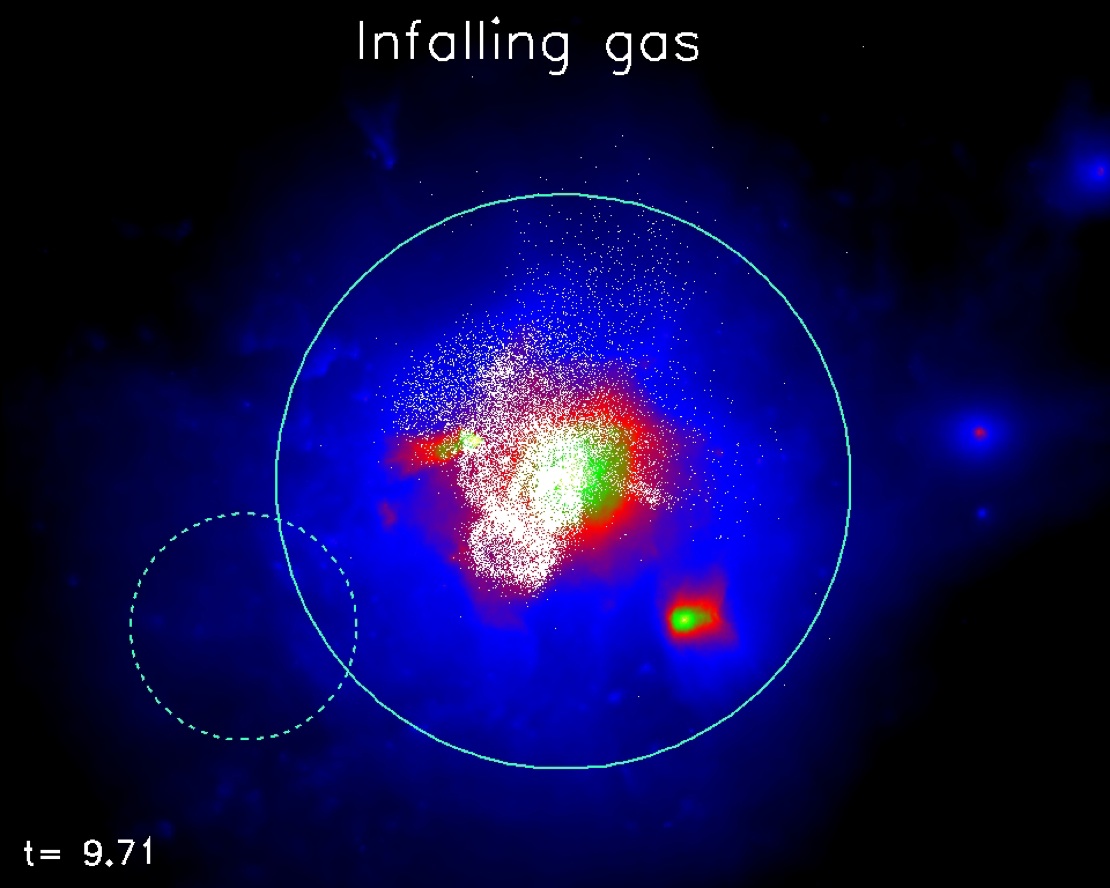}
\caption{ DM (left panels)  and ICM  (right panels) particles of the substructure are shown in false color. The different rows correspond to the time $t_c$ (central panel), 600 Myr before (upper panel) and 600 Myr after (lower panel). Two movies can be downloaded from http://www.astro.lsa.umich.edu/$\sim$rasia/REL\_SCA\_MERGER.php}.
\label{fig:detach}
\end{center}  
\end{figure} 

We summarize here only the basic concepts behind the simulation process;  further details can be found in \cite{dolag.etal.09}.
Both clusters are originally identified in the same dark-matter-only
cosmological simulation, 479 Mpc$h^{-1}$ on a side \citep{yoshida.etal.01}, and then
re-simulated using a more detailed treatment of the intracluster
medium physics. 
 The re-simulation technique used takes advantage
of the Zoomed initial condition method \citep{Tormen.etal.97} which
largely increase the spatial and mass resolution.  As result, the `1:1 merger' cluster has 2 million particles inside $R_{200}$ \footnote{$R_{200}$ and $R_{500}$ are defined as the radius of the sphere 
whose density is 200 and 500 times (respectively) the critical density of the universe at that redshift.}, while the `bullet-like' system has a factor of 20 more particles.  

The re-simulations employ GADGET-2 \citep{gadget2}, in which we introduced a
uniform and evolving UV background \citep{haardt&madau}, star formation from a multiphase
interstellar medium, a prescription for galactic winds triggered by SN
explosions \citep{springel&hernquist03} and the influence of 1/3-Spitzer thermal
conduction.  In the `bullet-like' simulation, we use the
method by \cite{Dolag.etal.05a} to limit the artificial viscosity typical of SPH simulations
 as well as the model of chemical enrichment by \cite{tornatore.etal.07}.
The cosmology is a flat 
$\Lambda CDM$ model with $\Omega_{0,m}=0.3$ for the present 
matter density parameter, and $\Omega_b=0.039$ for the baryonic density, leading to a baryon fraction of 0.13.
 The Hubble constant is $h=0.7$ in units 
of 100 km s$^{-1}$ Mpc$^{-1}$, and the rms level of density fluctuations 
within a top-hat sphere of 8 $h^{-1}$ Mpc radius is $\sigma_8=0.9$.

\paragraph{Bullet-like cluster.} This halo is the most massive system of the 25 investigated clusters of the original sample ($g8.a$ in Table 2 of Dolag et al. 2009).  It is in the central position of a very dense environment which includes four other systems of mass $M_{200} > 1 \times 10^{14} M_{\odot}$ within a 10 Mpc radius.  The halo is connected with its companions through several filaments from which it is accreting material.

In  Fig.~\ref{fig:detach}, we show the dynamics of both dark matter (left column) and gas (right column) components of the substructure before, during, and after the merger. 
The substructure moves from North-West to South-East.  In the top panels we show the moment ($t_{before} \sim$ 8.5 Gyr, $z=0.49$) when the two radii, $R_{500}$ (white circles), of the halo and of the secondary object intersect for the first time.  The central panels show the instant ($t_c \sim 9.1$ Gyr, $z=0.4$) when the centers of the two systems are coincident.  Finally, the lower panels show the situation ($t_{after} \sim$ 9.7 Gyr, $z=0.34$) when most of the mass of the substructure exits from the $R_{500}$ sphere of the main object. The temporal range from $t_{before}$ and $t_c$ and from $t_c$ and $t_{after}$ is 600 Myr. 

During the interaction, the two components are experiencing different forces and are influenced in a different way by the merger. The gas is slowed down during the infall and shows already an asymmetrical tail at time $t_{before}$. During the merger, ram-pressure stripping confines the plasma to a region close to the center of the main cluster, and some gas stays even after $t_{after}$. Since the dark matter feels only the gravitational forces, after the passage through the center it continues its track relatively unperturbed. The star particle clumps that represent the galaxies in the simulation (not shown in the Figures) follow very closely the dark matter particles behavior, lagging only slightly behind the dark matter at time $t_{after}$.

While the most evident feature of this figure is the separation of the baryonic plasma from the non-baryonic component, other  elements are distinguishable.  First, the geometry of the cluster is not spherically symmetric, but elongated in West-East direction. Second, note the presence of several other clumps which fall into the main halo.  Among them a chief role is played by a very cold and massive satellite which is located on the West side on the central panels and on the South-West on the lower images. In what follows, we denote as $t_{cold}$ the instant when this satellite enters within $R_{500}$ of the main  cluster.

To trace the interaction evolution we use 26 snapshots from redshift $\sim$ 0.5 (or Hubble time equal to $t = 8.2$ Gyr) to redshift 0.28 ($t=10.2$ Gyr). 
The time step between consecutive snapshots is generally 150 Myr, excluding the period from $t_c$ and $t_{after}$
where a shorter interval of 30 Myr is used to increase the temporal resolution.

\paragraph{1:1 merger case.} 
The second situation we follow is the final phase of the formation of a moderate mass halo with $M_{200}=5 \times 10^{14} M_{\odot}$ at $z=0$. As in the former case, this halo also lies in a dense environment including 5 objects with $M_{200}$ greater than $0.5 \times 10^{14}$ within 10 Mpc.  However, it is not the most massive member of this supercluster ($g1.b$ in Table 2 of Dolag et al. 2009)

\begin{figure}[h!]
\begin{center} 
\includegraphics[width=0.5\textwidth]{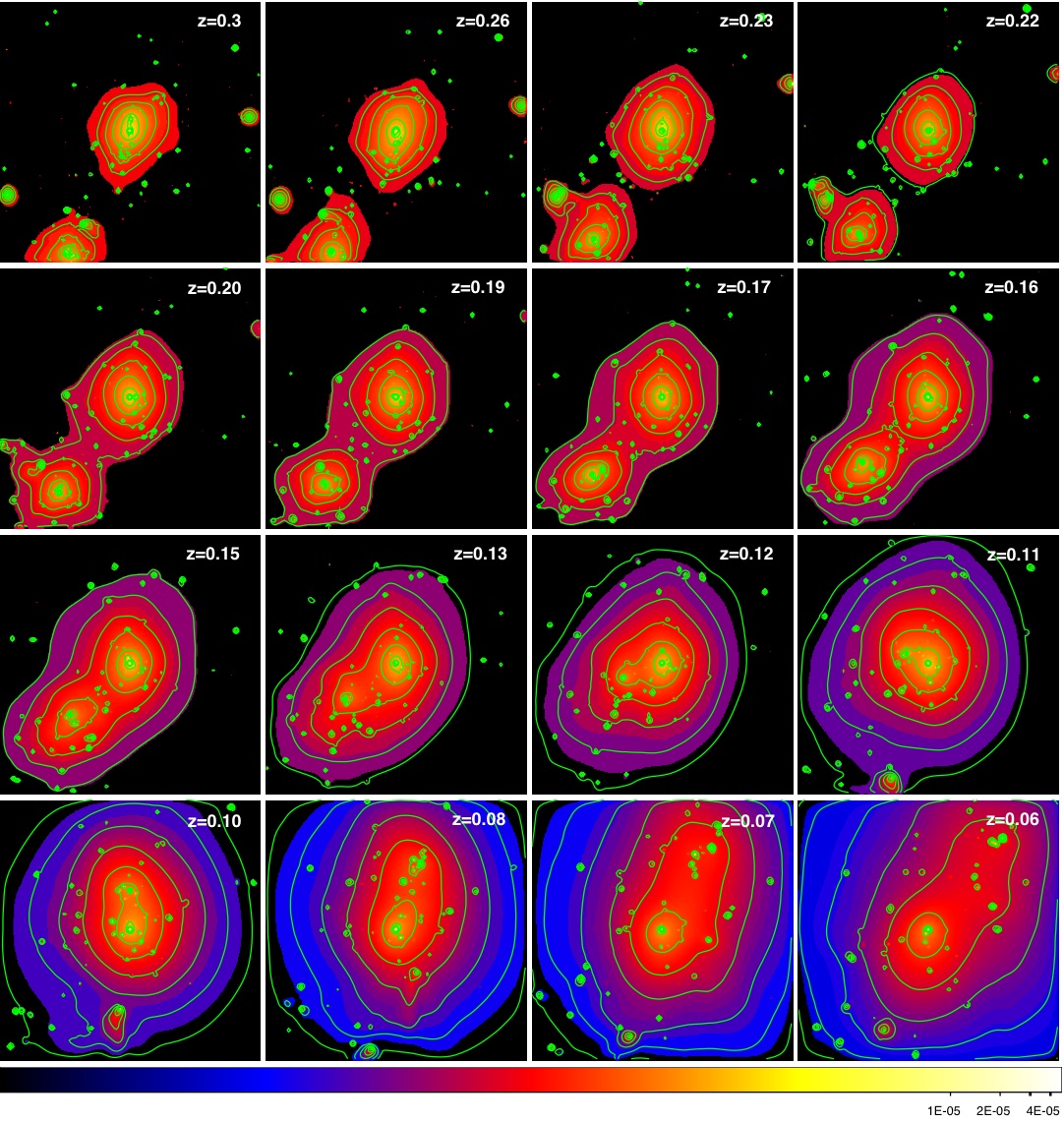}
\caption{ Intrinsic flux images of the `1:1 merger case' taken from redshift 0.3 to redshift 0.06.  The field of view is fixed to 4 Mpc.  
The color scale is the same for all panels and is in units of photons/s/cm$^2$. }
\label{fig:g1bflux}
\end{center}  
\end{figure} 

We study the most recent major merger involving two objects of comparable mass. In Fig.\ref{fig:g1bflux},  we show 16 intrinsic flux maps that exhibit the evolution of this merger. The lower mass system arrives from South-East and passes to the East side of the center.  It reaches a closest distance of 300 kpc at time $t_c$, and executes a spiral motion before being complete absorbed by the principal halo. The line of sight that we chose is that which minimizes the impact parameter  and might create the biggest mis-interpretation of the X-ray data. Indeed, the flux map correspondent to $z= 0.11$ in Fig.2 present regular  iso-flux contours, indication of a relaxed cluster, and only one bright spot.

For the rest of the paper, we consider 15 snapshots from redshift 0.3 ($t=10.07$ Gyr) to redshift 0.09 ($t=12.23 $Gyr), separated by 150 Myr.  As above, $t_{before}$ denotes the instant when the two $R_{500}$ are in contact.

\section{Intrinsic scaling evolution}
\label{sec:simres}

\begin{figure}
\begin{center} 
\includegraphics[width=0.5\textwidth]{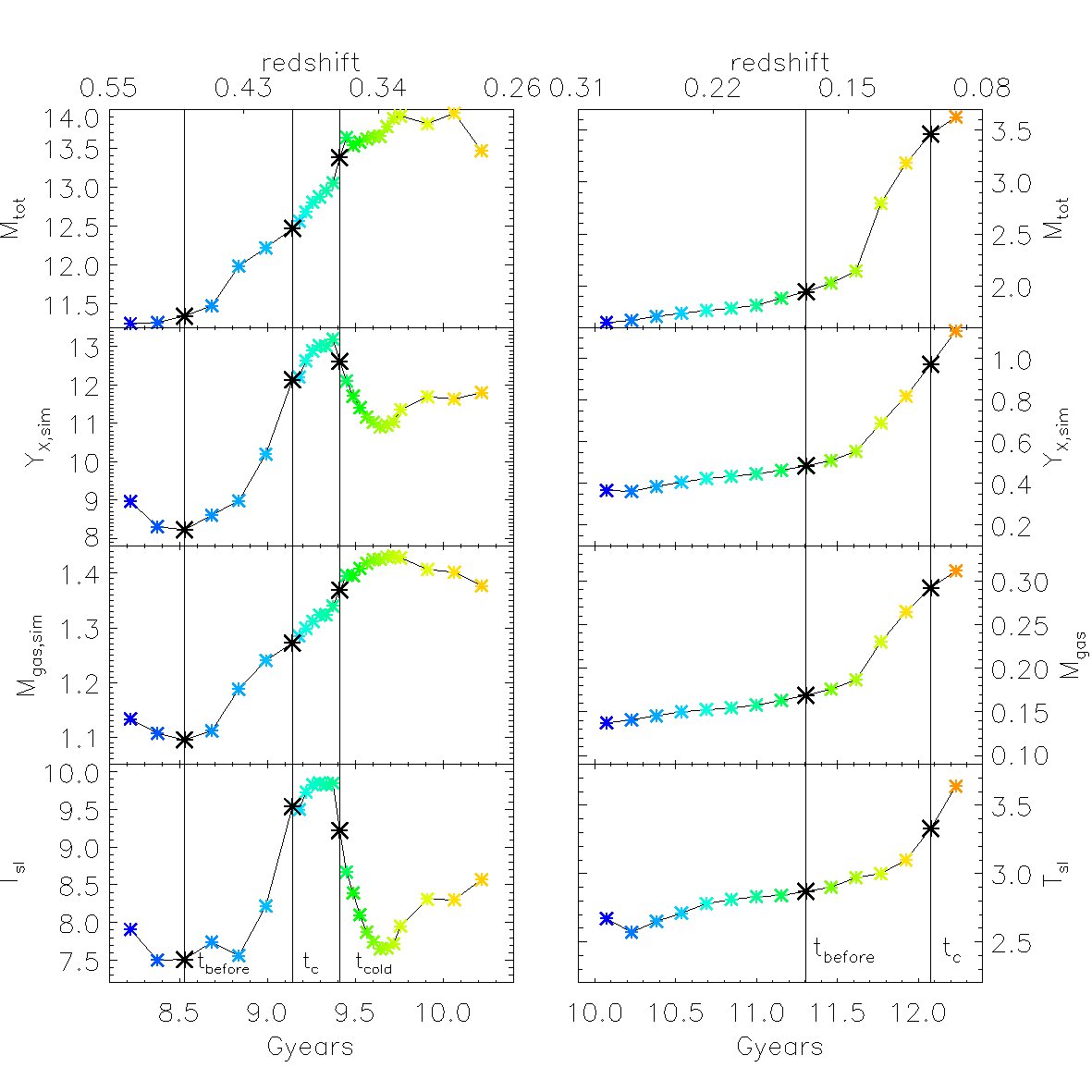}
\caption{Evolution of the spectroscopic-like temperature, $T_{sl} [keV]$, gas mass, $M_{gas,sim} [10^{14} M_{\odot}]$, $Y_{X,sim}= M_{gas,sim} \times T_{sl} [10^{14} M_{\odot} \times keV]$ parameter, and total mass, $M_{tot} [10^{14} M_{\odot}]$ (bottom to top panels).  On the left side we present the `bullet-like' case, while on the right side the `1:1 merger' case. The vertical lines indicate important moments in the clusters' evolution: $t_{before}$, $t_c$, and $t_{cold}$ (see text in Sec.\ref{sec:simul} for details)}
\label{fig:evol}
\end{center}  
\end{figure}


\begin{table}
\caption{Parameters of the fitting formula $M_{14,tot}*E(z)^{\beta}=M_0 (X/x_0)^{\alpha}$ in the case of all redshifts and all clusters \cite[Tab. 2 of][]{kravtsov.etal.06}. $M_0$ is changed accordingly to the different baryon fraction}
 \begin{tabular}{l c c c c c}  
 & & \\ 
 & $M_0$ & $\alpha$ & $x_0$ & scatter & $\beta$ \\ 
\hline \\  
$T $[keV]			&2.55 & 1.521  & 3     & 19.5    &  1 	\\ 
$M_{gas} [10^{14}M_{\odot}]$   	&2.42 &  0.921 & 0.2  & 10.7    &   0 	\\ 
$Y_X [10^{14}M_{\odot} \times$ keV$]$  		&1.95 &  0.581 & 0.4  &   7.1    &   2/5	\\ 
 
\end{tabular}  
\label{tab:rel}
\end{table}

For each snapshot, we locate the center of the halo as the minimum of the potential well. We compute the gas mass,
$M_{gas, sim}$, summing the gas particles inside $R_{500}$, and the total mass, $M_{tot}$, considering the contribution by all species of particles (dark matter, gas, and stars) . 
The choice of this radius is motivated by both theoretical \citep{evrard.etal.96,rowley.etal.04, nagai.etal.07} and observational works \citep{vikh.etal.06} that establish this as a scale within with scaling relation scatter is minimized.

To identify the temperature, we apply the spectroscopic-like definition, $T_{sl}=  \sum W T dV/ \sum W dV$ with $T$ the temperature of each gas particle, $n$ its density and $W$ equal to $n^2/T^{(0.75)}$  \citep{tsl}. The sum is extended to all the particles with temperature larger than 0.5 keV within [0.15 1] $R_{500}$.  The exclusion of the inner region is conventional among observers and simulators since it guarantees that X-ray results are not influenced by the activity of the cluster core and intrinsic analysis of simulations are not affected by the overcooling problem.
Finally, $Y_{X, sim}$ is the product of the previously measured quantities $Y_{X, sim}=M_{gas, sim} \times T_{sl}$. 

In Fig.~\ref{fig:evol} we show the time evolution of the total mass, the $Y_{X,sim}$ parameter, the gas mass, and the spectroscopic-like temperature of both halos. 
The evolutionary tracks of the masses of the bullet-like halo show several bumps which testify to the continuous accretion of material due to the infall of satellite objects (see also, Fig.~\ref{fig:detach}). Notice that many of these show a mass ratio which is higher than that we are focusing on in this work (1:10). Nevertheless,  this particular merger is the only one that produces an increase on temperature of 40\% and a Mach number of 2.5. After $t_{cold}$, the temperature has a rapid drop due to the arrival of the second, colder satellite. The decline is shallower for $Y_{X,sim}$ due to the combined effect of the increase of $M_{gas, sim}$.

In the case of the `1:1 merger' the evolutionary tracks are smoother and they all present only one big bump of order of 100\% for the masses and 40\% for the temperature due to the merger. In both cases, the increase in the temperature is not simultaneous with the increase in the mass, a delay of few hundred mega years is needed to allow shocks to develop and heat the intra-cluster medium.


\section{X-ray-observed scaling evolution}
\label{sec:xray}

\begin{figure*}
\begin{center} 
\includegraphics[width=0.45\textwidth]{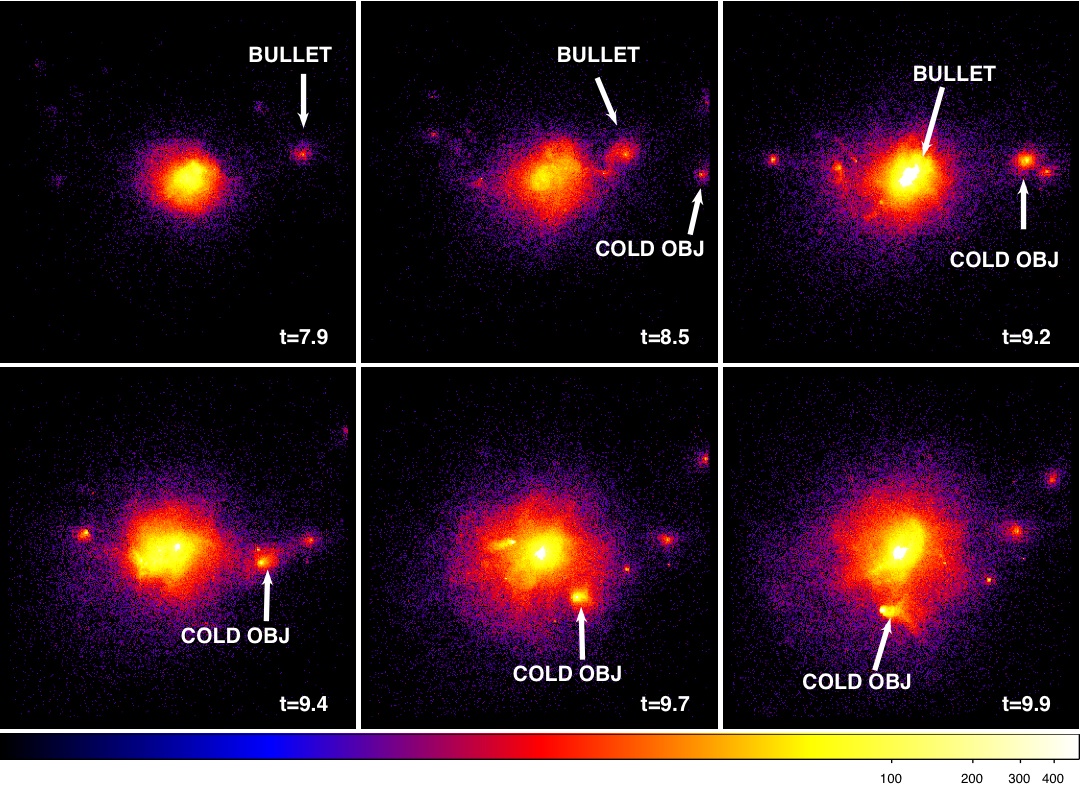}
\includegraphics[width=0.45\textwidth]{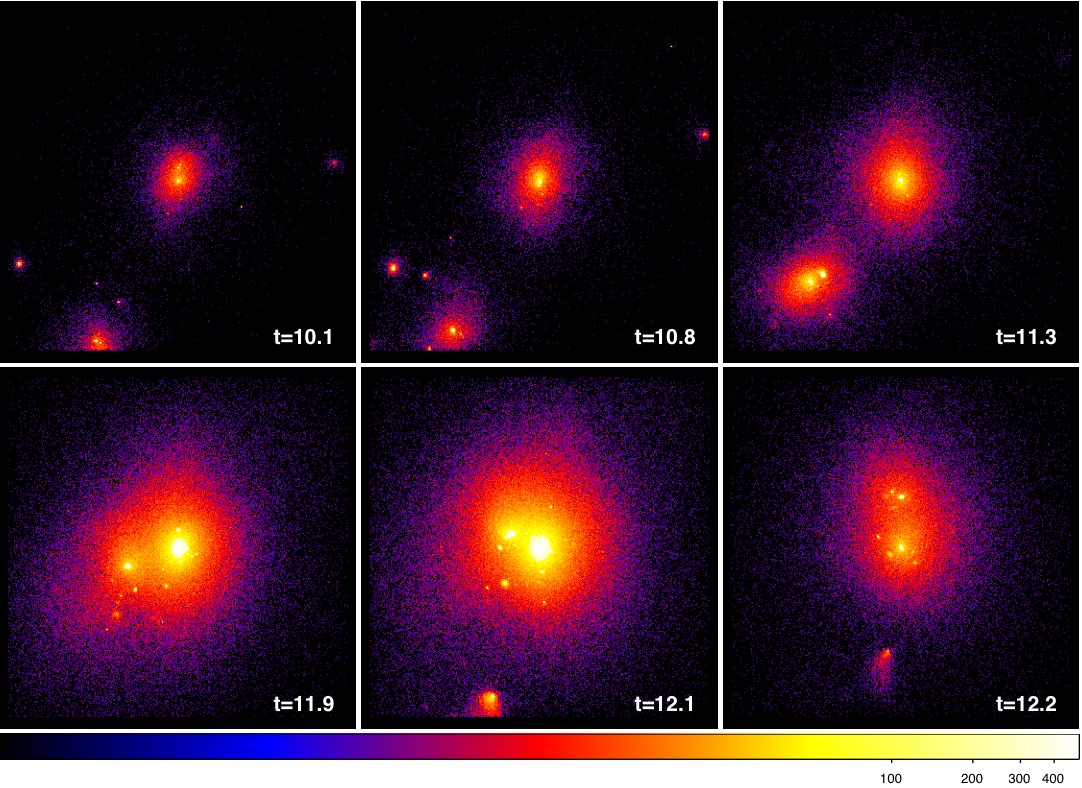}
\caption{Photon images in the [0.7 2] keV soft band of `bullet-like' (left panels) and `1:1 merger' cases (right panels). The field of view is 16 arcmin and the exposure time 1 Msec for each panel. The entire collection of X-ray images can be found at http://www.astro.lsa.umich.edu/$\sim$rasia/ANALYSIS\_bullet.php and http://www.astro.lsa.umich.edu/$\sim$rasia/ANALYSIS\_1\_1.php}
\label{fig:soft_img}
\end{center}  
\end{figure*}

We create X-ray images for each snapshot with our X-ray MAp Simulator ({\em X-MAS}). The characteristics of this software package are described in detail in other works  \citep{xmas,rasia.etal.06,rasia.etal.08}. To create the photon event file, we 
 use the ARF and RMF of the Chandra ACIS-S3 detector aimpoint. We fix the redshift at the value of the simulated time frame and the equivalent hydrogen column density to $n_H=10^{20} cm^{-2}$. In the `1:1 merger case', the metallicity is one third of  the solar abundances by \cite{anders&grevesse}. In the `bullet-like' halo, instead, we use the detailed star formation history model by \cite{tornatore.etal.07} and each particle has different abundances for $C$, $N$, $O$, $Mg$, $Si$, $Fe$ and a constant value for the remaining elements \citep[see also][]{rasia.etal.08}. The field of view of our images is 16 arcmin and the depth we use in the projection is 10 Mpc. The exposure time is 1 Msec.

Selected photon images in the soft band, [0.7 2] keV, are presented in Fig.~\ref{fig:soft_img} for both clusters.
In the images of the bullet-like cluster, the position of the bullet and of the cold object discussed in Sec.~\ref{sec:simul} are evident. As in Fig.~\ref{fig:evol}, we note also the presence of many other clumps with comparable size to the bullet. All these structures play 
an important role in perturbing and modifying the ICM appearance. The `1:1 merger' cluster also shows the existence of small blobs, but they are much less massive than the merging object we are considering.

For each time step, we apply the following X-ray spectral image processing to obtain the gas mass $M_{gas, X}$ , the temperature $T_X$ and the parameter $Y_X$. It is important to stress that we proceed on every single time step in an independent way, i.e. without knowledge either of the information stored in the simulation itself or of its past history or future evolution.


\paragraph{Surface Brightness.}
Using {{\em CIAO}} tool \citep{fruscione.etal.06} we extract soft band  images in the [0.7 2] keV band per each snapshot and use them to identify cold dense regions with the wavelet algorithm of \cite{vikh.etal.98}. These regions are masked and excluded from all the analysis. We proceed to extract the surface brightness profiles in 30-40 linearly spaced annuli.   For the center, we generally use the centroid of the external and more spherical iso-flux contours.
The profiles, stored as counts versus radius, span over a radial range from $\sim 0.1 \times R_{500}$  to $\sim 2 \times R_{500}$.

\paragraph{Temperature.}
To measure the X-ray temperature, we extract the spectrum in the [0.5-7] keV band of the region inside $R_{500}$ with the exclusion of the masked regions and of the inner 15\% of $R_{500}$. At a first pass, $R_{500}$ is defined directly from the simulation (see Sec.\ref{sec:simul}). In a second step, we derive $R_{500}$ 
following the X-ray procedure described below.  We refer to this measure as $R_{500,X}$. The X-ray spectra are fitted using a  $\chi^2$ statistic in the XSPEC package \citep{arnaud96} with a single temperature  MEKAL model by fixing input values of the hydrogen column density ($n_H=10^{20} cm^{-2}$), redshift and, in the '1:1 merger case', the metallicity ($Z=0.3 Z_{\odot}$).

\paragraph{Gas Mass.}

Following \cite{vikh.etal.06}'s procedure, we fit simultaneously the surface brightness and the temperature profiles using the following expressions:

\begin{eqnarray}
n_p n_e &=& n^2  \frac{(r/r_{c})^{-\alpha_n}}{[1+(r/r_{c_1})^2]^{3\beta-\alpha_n/2}}  \frac{1}{[1+(r/r_s)^{\gamma}]^{\epsilon/\gamma} } 
\end{eqnarray}
\begin{equation}
T=T_0 \frac{(r/r_t)^{-a}}{[1+(r/r_t)^b]^{c/b}},
\end{equation}
 The surface brightness fitting formula is an extension of the $\beta$ model \citep{cavaliere&ff76} with the addition of an inner power-law cusp (the $\alpha$ term), and a change in the slope in the external region (slope modified by $\epsilon $ around $r_s$). 
The temperature expression is the product of a broken power law.  A more comprehensive description includes additional terms that describe the inner core region  \citep[see][]{vikh.etal.06}, but we do not consider them here since we exclude the inner 15\% of $R_{500}$  from our profile analysis.

\paragraph{R$_{500,X}$.}
We derive an estimate of the scale radius, $R_{500, X}$, using only the X-ray information.  This iterative method \citep{kravtsov.etal.06} is based on knowledge of the $M-Y_X$ scaling relation, of the gas mass profile, $M_{gas}$, and of different measurements of the temperature, $T_{X}$, inside subsequent estimate of $R_{500, X}$. Calling the first guess OLD, the equation we need to solve to find the NEW $R_{500, X}$ is based on the two definitions of $M_{tot}$, 
\begin{eqnarray}
\frac{4\pi}{3}R_{500,X, NEW}^3 \rho_{cr}  &=&  M_0 E(z)^{\beta} (Y_{X,OLD}/x_0)^{\alpha} \nn 
              &=&M_0 E(z)^{\beta} [(M_{gas,OLD} \times T_{X, OLD})/x_0]^{\alpha},
\end{eqnarray}
where $\rho_{cr}$ is the critical density at the considered redshift, $z$; $E(z) = H(z)/H_0 = [\Omega_M(1+z)^3 +  \Omega_{\Lambda}]^{0.5} $ \citep{peebles93} is the evolution factor; $M_0$, $x_0$, $\alpha$ and $\beta$ are the specific constants of the $M-Y_X$ relation (see Table~\ref{tab:rel}). $Y_{X,OLD}$, $M_{gas,OLD}$, and $T_{X,OLD}$ are the quantities measured inside the previous attempt of $R_{500,OLD}$.

This equation defines the next value, $R_{500, X, NEW}$, for which we repeat the measurements of temperature and the calculation of gas mass.  The process is iteratively repeated until subsequent estimates of $R_{500}$ are consistent within 3\%.  In our sample, we begin our calculation assuming the radius computed from the simulation directly, $R_{500, sim}$. For this reason, in 75\% of the cases we stop the iteration after the first round.

\section{Results}

\begin{figure}
\begin{center} 
\includegraphics[width=0.45\textwidth]{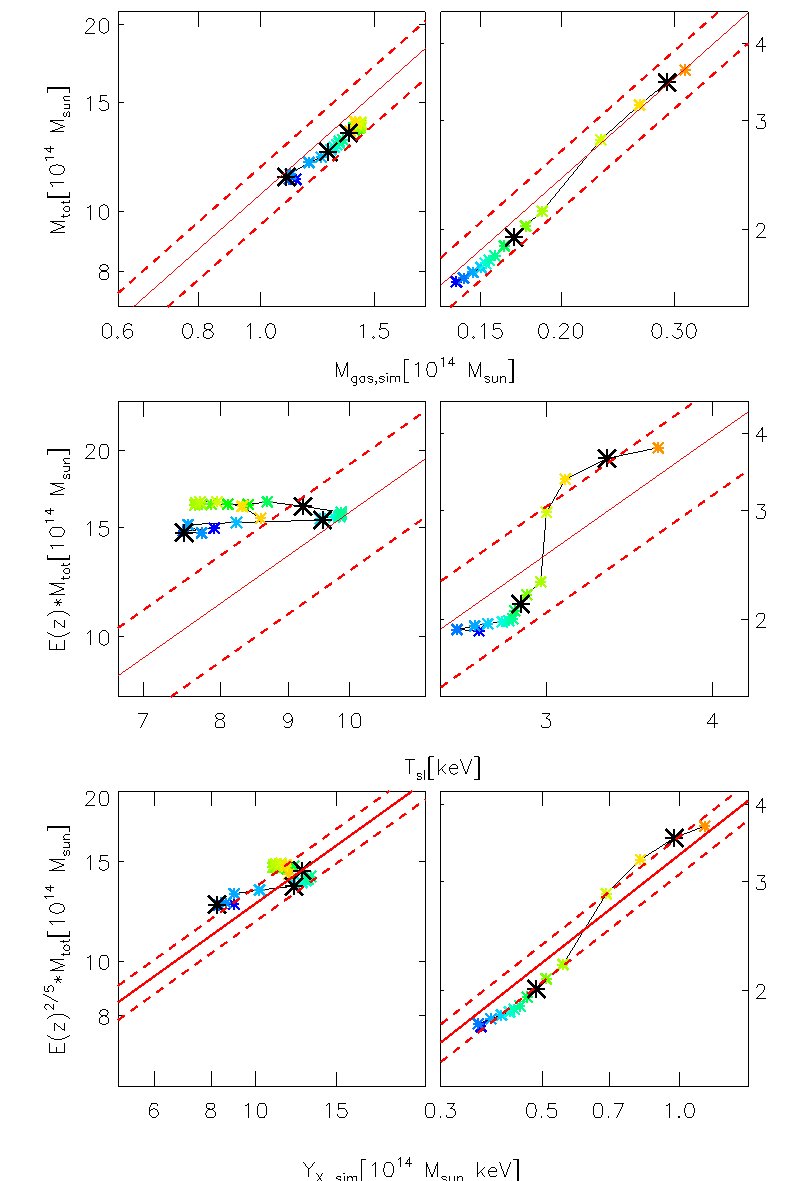}
\caption{Scaling relations from intrinsic analysis of the simulated halo. Left panels refer to the bullet-like halo, while right panels to the 1:1 merger case. From the top, we show $M_{tot}-M_{gas,sim}, M_{tot}-T_{sl}, and M_{tot}-Y_{X,sim}$. The total mass is multiply by $E(z)$ to a power equal to 0, 1, 2/5, respectively (see text for the definition of $E(z)$). The different color are the same of Fig.\ref{fig:evol}. Red solid and dashed lines represent Kravtsov et al. (2006) relation and its scatter (see Table.~\ref{tab:diff}).}
\label{fig:relsca_sim}
\end{center}  
\end{figure} 

\begin{figure}
\begin{center} 
\includegraphics[width=0.45\textwidth]{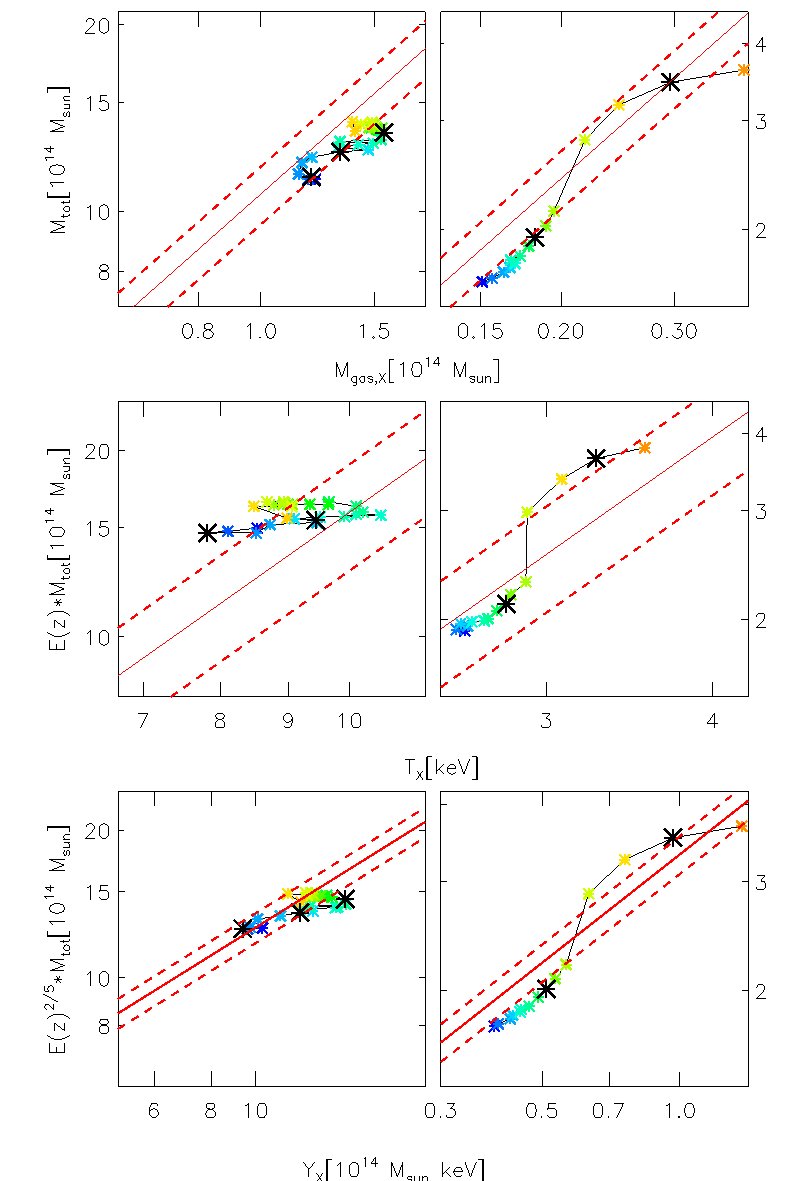}
\caption{The same as Fig.~\ref{fig:relsca_sim} but with the measurement of $M_{gas,sim}$, $T_{sl}$, and $Y_{X,sim}$ done following the X-ray procedure.}
\label{fig:relsca_x}
\end{center}  
\end{figure}

The scaling relations results for the quantities intrinsically computed in the simulation (Sec.\ref{sec:simres}) and the measurements done following the X-ray procedure (Sec.~\ref{sec:xray}) are presented in Fig.~\ref{fig:relsca_sim} and Fig.~\ref{fig:relsca_x}, respectively.  
In both figures, the $y$-axis is to the total mass multiplied by the evolution factor, $E(z)$ at the appropriate $\beta$ power (see Table 1).  Evolution powers are $\alpha=0$ for gas mass, $\alpha=1$ for  temperature,  and $\alpha=2/5$ for$Y_{X}$. The color scheme for points in Figs.~\ref{fig:relsca_sim} and \ref{fig:relsca_x} is the same used in Fig.\ref{fig:evol}, with the black symbols specifying the significant moments, $t_{before}$,  $t_c$, and, in the case of the `bullet-like' cluster, $t_{cold}$.

In every panel, the solid red line shows the \cite{kravtsov.etal.06} scaling relation rescaled by the mean baryon fractions of our respective simulations.  We assume a baryon fraction $\Omega_b=0.039$ (see Sec.\ref{sec:simul}) while they assume $\Omega_b=0.04286$. The ratio, $1.099$, corrects $M_{gas}$. Considering that their average  $<M_{gas}/M_{tot}>$ was $\sim 0.09$, we correct also $M_{tot}$ by 1.008. Finally, the dashed lines represent their estimate of the $1\sigma$ scatter in each scaling relation. Tab.\ref{tab:rel} lists parameters of the scaling relations shown, rescaled for our cosmology.

\subsection{Intrinsic scaling relations}

The two systems under study have substantially different evolutionary behavior; the `bullet-like' cluster has a complicated history with the multiple accretion events while the `1:1 cluster'  evolves more smoothly.  The BL variation of mass and temperature are not synchronized, so there are several loops in the $M_{tot}-T_{sl}$ relation.  Moreover, the presence of the secondary, cold accretion causes significant movement in the $M_{tot}-T_{sl}$ plane toward the cool side of the mean relation (transition between the cyan points before $t_{cold}$ and the green points after $t_{cold}$). This variation is also reflected in the $M_{tot}-Y_{X, sim}$ relation.

The scaling evolution of the 1:1 case is more regular.  Except for a phase delay during the merger itself, the gas mass, temperature, and $Y_{X, sim}$ observables (top to bottom) in Fig.~\ref{fig:relsca_sim} increase in a manner that largely reflects the appropriately scaled growth in total mass.   The temperature and $Y_X,sim$ parameter experience delays of $\sim 150$~Myr (see also Fig.~\ref{fig:evol}), as shocks expand and heat the plasma. That produces jumps in the $M_{tot}-T_{sl}$ and $M_{tot}-Y_{X, sim}$ relations that changes the sign of the mass proxy residuals, from slightly underestimating to overestimating the true mass.  Yet the entire evolutionary histories of the 1:1 halo's scaling relations lie within the 1-$\sigma$ region proposed by \cite{kravtsov.etal.06}.  Since their analysis was based on measures derived from synthetic X-ray observations, it is not surprising that our intrinsic data should show smaller scatter.  We turn now to our estimates of X-ray observable quantities.

\subsection{X-ray observed scaling relations}

The scaling relation  between the total mass and the gas mass in Fig.\ref{fig:relsca_x} shows a similar behavior of Fig.\ref{fig:relsca_sim}: the two systems move along the \cite{kravtsov.etal.06} relation without major deviation caused by the merging. In general, the X-ray gas mass estimates are slightly larger than the true gas mass, by $5.5\%$ on average.  This value agrees with the results of \cite{nagai.etal.07}, who found a bias of $5.8\% \pm 5\%$ for their unrelaxed systems at redshift zero.
The X-ray temperatures are hotter for the bullet halo in the final phase of the merger and, in particular, after $t_{cold}$ (green points).  The temperatures of the 1:1 case track the intrinsic evolution with a high degree of fidelity.
Finally,  the combined behavior of the gas mass and temperature lead to $Y_{X}$ being slightly higher in the X-ray analysis.

\subsection{{\it A priori} difference between the two procedures.}

The differences between the intrinsic and X-ray  scaling relations are slight, and some of the difference can be explained by 
bias in setting the region within the quantities are measured.  
In this section, we examine whether and how the results are affected by the choice of the center, the radius and  the masking.
\paragraph{Center.}
In the intrinsic analysis of the simulated halos we define the center as the minimum of the potential well  
while in the X-ray analysis the center coincides with the X-ray centroid.
The two definitions agree well for the 1:1 case, but they often differ in the BL case, by typical values of 90 kpc.
Such an offset introduces a negligible effect in the temperature calculation.   While it might change the inner profile of the surface brightness, and thus modify the value of the gas mass, 
our analysis masks the innermost regions 
where this effect would be relevant.
 To explore the importance of the center definition, we test whether there is a correlation between the centers' offset and the gas mass difference
and found only a negligible positive correlation ( Pearson correlation coefficient equal to 0.3). We conclude that the gas mass difference cannot be explained by the different selection of the center.

\begin{figure}
\begin{center} 
\includegraphics[width=0.5\textwidth]{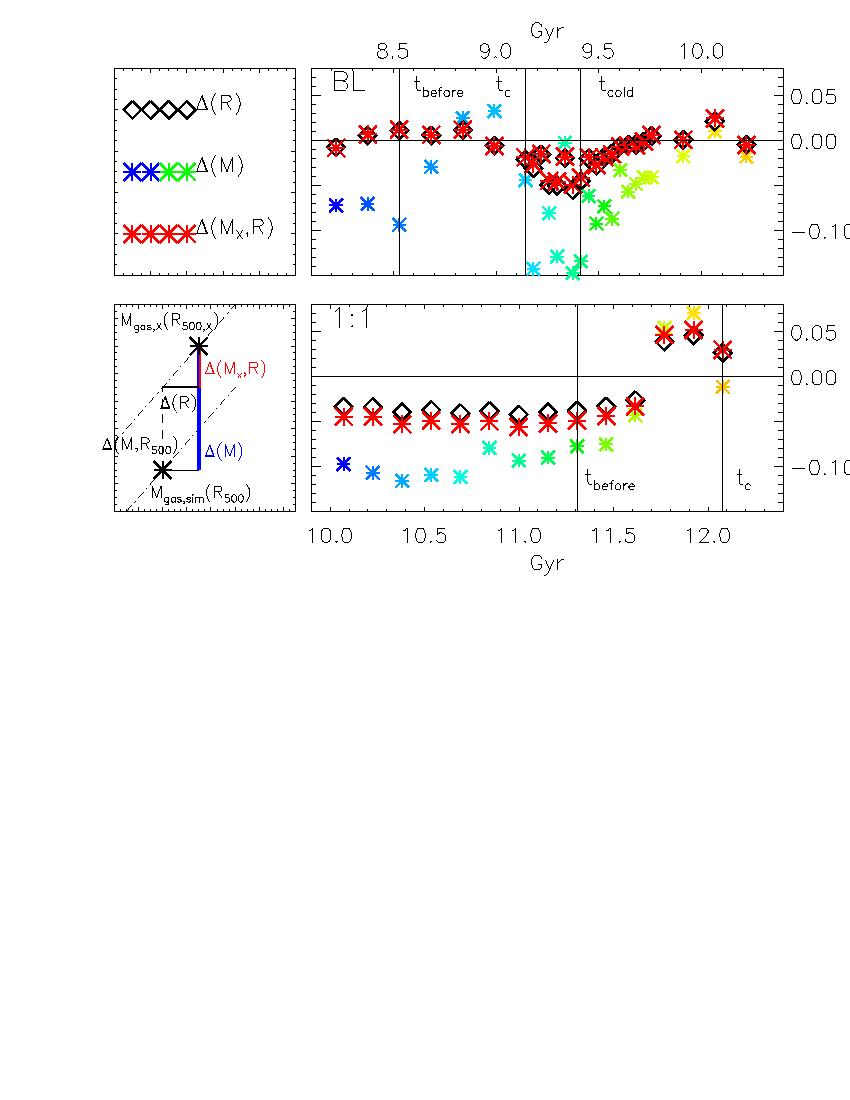}
\caption{$\Delta_M$ in colored points, $\Delta_R$ is black diamonds and $\Delta_{M_{X,R}}$ in red. See text for the definition of the various $\Delta$.}
\label{fig:compa}
\end{center}  
\end{figure}


\begin{table}[!h]
\caption{Percentage deviation between the true radius and the X-ray radius, $| \Delta_R|$,  the intrinsic gas mass and the X-ray one, $|\Delta_M|$; the X-ray gas mass at  the true radius and at the X-ray radius, $|\Delta_{M_X,R}|$; the X-ray gas mass and the intrinsic one at the true radius, $|\Delta_{M,R_{500}}|$; and the correlation between $\Delta_R$ and $\Delta_M$.}
 \begin{tabular}{l c c}  
 & & \\ 
                                           & BL & 1:1 \\ 
\hline \\  
$|\Delta (R)|$&                     1.8\%                      & 3.7\% \\
$|\Delta (M)|$   &     6.2\%                     & 8.1\% \\
$|\Delta (M_X,R)|$ &     1.7\%                      & 4.7\% \\
$|\Delta (M,R_{500})|$ & 4.9\% & 3.7\%\\
correlation$(\Delta_R,\Delta_M)$             &     0.70                         &0.95 \\

\end{tabular}  
\label{tab:diff}
\end{table}

\paragraph{Radius.}
On average, the true radius and the X-ray one are close. The maximum radii deviation,  $\Delta_R=(R_{500}-R_{500,X})/R_{500}$, is driven by the merger and it is always below 5\%. However, the sign of the deviations differ:  the X-ray radius is greater than the true one after the merger for the `bullet-like halo' , while it is the opposite for the `1:1 merger case'. We remind that in the $R_{500,X}$ computation, the 3\% requisite for further iterations is applied to two subsequent radius evaluations (see Sec. 4), the final difference with the intrinsic radius might, therefore, be higher. 

The small radial discrepancy directly affects the X-ray gas mass estimate since, at $R_{500}$, the gas density profile has a slope close to $-2$, implying that the gas mass is proportional to $R$ \citep{vikh.etal.06}.  A deviation of few percent in the radius implies a similar fractional error in  gas mass. 

In the bottom left panel of Fig.\ref{fig:compa}, we sketch the situation. The two dashed-dotted lines represent the intrinsic (lower) and estimated X-ray (upper) gas mass profiles.
The two large points indicate the value of the two gas mass at the two respective radii. The difference between the two radii,  $\Delta (R)$ (the two horizontal lines), is similar to the difference of the X-ray gas mass computed at the two radii, $\Delta (M_X,R)$ (short red vertical line at the right). For our halos, the situation is outlined in Table~\ref{tab:diff} and on the right panels of Fig.\ref{fig:compa}: the $\Delta (R)$ shifts (black diamonds) almost always coincide with the $\Delta (M_X,R)$ (red crosses).  
The total mass difference, $\Delta (M)$ (the long blue vertical line at the right in the bottom left panel) is represented by colored points in the left panels. In addition to $\Delta(M_X,R)$ it includes also the discrepancy between the X-ray and the intrinsic gas mass computed at a fixed radius (dashed vertical line on the bottom left, $\Delta (M,R_{500})$).  The last difference could be caused either by too restrictive assumptions on the geometry of the systems or by the presence/exclusion of the cold blobs. Even if these effects are hardly controllable, we can still reduce the systematics by computing correctly the radius. This will decrease the disagreement between the two masses  from 6.2\% to 4.9\% for the BL and from 8.1\% to 4.7\% for 1:1 case. 
 Notice that the X-ray radius is both an under- and over-estimate of the true value. However, the implication on the gas mass are larger in the over-estimate scenario. As consequence a wrong X-ray radius derivation does not simply increase the scatter but skews the gas mass distribution, changing, de facto, the scaling relation.

 These variations in mass are consistent  with the results of \cite{meneghetti.etal.09}. Using simulations from the our sample \citep{dolag.etal.09} they found a difference in gas mass of $7\% \pm 3\%$ when computed at the same lensing radius.

The $Y_X$-based scaling technique to compute the mass and radius, described in Sec.~\ref{sec:xray}, has inherent scatter that drives systematic error.  If the cluster is naturally below the $M_{tot}-Y_X$ scaling relation (such as the 1:1 halo at the beginning), the X-ray radius inferred through Eq.~3 will be greater than the true one. As consequence the cluster size will be shifted towards larger values in both the $M_{gas}$ and $Y_X$ measures.  
 The increase of the scatter along these proxies-axes will be the same, however, the total mass scatter will be smaller for $Y_X$  which has a lower power law.

\paragraph{Masked regions}
Another source of difference from the procedures  followed in Sec.~\ref{sec:simres} and Sec.~\ref{sec:xray} is the masking technique.  In the X-ray analysis we use a wavelet algorithm to exclude localized peaks present in the intra-cluster medium, and to compute the spectroscopic-like temperature we cut all the particles below 0.5 keV independently on the particle density or position. 
 Masking the cold regions is a fundamental step to link our work to the observed X-ray clusters. With this procedure we ensure that (i) the results based on the mock images are not influenced by the presence of the overcooling problem affecting radiative simulation \citep{borgani&kravtsov}; (ii) the appearance of the remaining X-ray emission is more similar to real X-ray observations \citep{boehringer.etal.10}; (iii) the X-ray analysis we perform mirrors observers approach.  

The measurement most affected by the masking is the temperature, as comparison of Fig.\ref{fig:relsca_sim} and Fig.\ref{fig:relsca_x} attests.  In the X-ray analysis, we exclude the secondary cold satellite.  As a consequence, the X-ray temperature between $t_c$ and $t_{cold}$ (green points) does not drop as dramatically as the intrinsic spectroscopic-like one, and the mass-temperature scatter is therefore lower. 

\section{Conclusion}
We study the behavior of X-ray proxies for total mass (gas mass, temperature, and their product, $Y_X$) in case of two extreme merger events.  The evolution of intrinsic, three-dimensional measures is compared to estimates derived from projected, X-ray spectral maps with high signal-to-noise.  We create photon event files using our X-ray MAp Simulator and analyze them with standard procedures.

Specific behaviors of the scaling relations, and lessons learned from comparing intrinsic measures with those determined from synthetic observations, are:

\begin{itemize}
\item {\bf Code comparison.} After a small correction for cosmic baryon fraction differences, the scaling relation behavior of both merger scenarios lies within the $\pm 2\sigma$ regions expected from ART simulations \citep{kravtsov.etal.06}.  This agreement in non-trivial, given our use of a different code (GADGET) employing a somewhat different physical treatment.   
\item {\bf Gas mass.} The intrinsic gas mass is a very good proxy for the total mass, growing almost contemporaneously and linearly with it in both merger cases.  The X-ray gas mass estimate is biased high by typically 5\%, due to asphericity in the hot ICM that is not accounted for in the model. The gas mass measurement is not affected by the masking technique but strongly depends on possible bias in the radius definition (see below).
\item {\bf Temperature.} The temperature is the X-ray proxy that shows the largest scatter and is the most influenced by the presence of substructures in the intra-cluster medium. Masking cold satellite emission reduces the scatter substantially.
\item {\bf X-ray SZ Estimate, $Y_{X}$.} The scatter in the mass$-Y_{X}$ relation is maintained low during the  both entire merging processes. 
\item {\bf Center.} The cluster center is defined independently in the intrinsic and X-ray analysis, being the minimum of the potential well in the former and the X-ray centroid in the latter.  Mis-centering differences up to $\sim$ 100 kpc are present for the BL cluster, but they do not significantly influence the mass proxy estimates at large radius.  
\item {\bf Radius.}   Estimates of the radial scale, $R_{500}$, do affect mass proxy estimates, with typical error $\sim 5\%$ in both merger cases .  The X-ray iterative derivation of the radius might conduce to an over-estimate with reflections on the gas mass measurements. The total mass scatter of the $M_{tot}-M_{gas}$ scaling relation will be more affected than that of $M_{tot}-Y_X$ due to the different power laws.
\item {\bf Masking.} The masking procedure plays a critical role when comparing radiative simulations with X-ray observations. Results from simulation that do not exclude any regions can overestimate temperature deviation with respect to the mean. We show that this has a large impact on the temperature calculation but not on the gas mass estimate.
\end{itemize}

The results of this paper, combined with the finding by \cite{yang.etal.09} that the scatter is lower at higher redshift even considering merging clusters, are encouraging for future missions (such as e-ROSITA\footnote{http://www.mpe.mpg.de/heg/www/Projects/EROSITA/main.html}  or the Wide-Field X-ray Telescope\footnote{http://wfxt.pha.jhu.edu/}) that aim to observe clusters at high redshift, where more frequent mergers are  expected.


\acknowledgements{ We would like to thank Stefano Borgani, Stefano Ettori, Gabriel Pratt, Alexey Vikhlinin and the anonymous referee for their constructive comments.
We acknoweldge
  financial contribution from contracts ASI-INAF I/023/05/0, ASI-INAF
  I/088/06/0 and INFN PD51. Support for this work was provided also by NASA through {\it Chandra} Postdoctoral Fellowship grant number PF5-70042 awarded by the {\it Chandra} X-ray Center, which is operated by the Smithsonian Astrophysical Observatory for NASA under the contract NAS8-03060. ER acknoweldges the Michigan Society of Fellow. AEE acknowledges support from NSF AST-0708150 and NASA ATP Grant NNX10AF61G. KD acknowledges the support from the DFG Priority Programme 1177 and additional support by
the DFG Cluster of Excellence "Origin and Structure of the Universe". The work has been performed under the HPC-EUROPA2 project (project number: 228398) with the support of the European Commission - Capacities Area - Research Infrastructures. }

\bibliographystyle{apj}

\begin{thebibliography}{58}
\expandafter\ifx\csname natexlab\endcsname\relax\def\natexlab#1{#1}\fi

\bibitem[{{Anders} \& {Grevesse}(1989)}]{anders&grevesse}
{Anders}, E. \& {Grevesse}, N. 1989, \gca, 53, 197

\bibitem[{{Arnaud}(1996)}]{arnaud96}
{Arnaud}, K.~A. 1996, in ASP Conf. Ser. 101: Astronomical Data Analysis
  Software and Systems V, 17--+

\bibitem[{{Arnaud} \& {Evrard}(1999)}]{arnaud&evrard}
{Arnaud}, M. \& {Evrard}, A.~E. 1999, \mnras, 305, 631

\bibitem[{{Biviano} {et~al.}(2006){Biviano}, {Murante}, {Borgani}, {Diaferio},
  {Dolag}, \& {Girardi}}]{biviano.etal.06}
{Biviano}, A., {Murante}, G., {Borgani}, S., {Diaferio}, A., {Dolag}, K., \&
  {Girardi}, M. 2006, \aap, 456, 23

\bibitem[{{B{\"o}hringer} {et~al.}(2010){B{\"o}hringer}, {Pratt}, {Arnaud},
  {Borgani}, {Croston}, {Ponman}, {Ameglio}, {Temple}, \&
  {Dolag}}]{boehringer.etal.10}
{B{\"o}hringer}, H., {Pratt}, G.~W., {Arnaud}, M., {Borgani}, S., {Croston},
  J.~H., {Ponman}, T.~J., {Ameglio}, S., {Temple}, R.~F., \& {Dolag}, K. 2010,
  \aap, 514, A32+

\bibitem[{{Borgani} \& {Kravtsov}(2009)}]{borgani&kravtsov}
{Borgani}, S. \& {Kravtsov}, A. 2009, ArXiv e-prints

\bibitem[{{Brada{\v c}} {et~al.}(2008){Brada{\v c}}, {Allen}, {Treu},
  {Ebeling}, {Massey}, {Morris}, {von der Linden}, \&
  {Applegate}}]{bradac.etal.08}
{Brada{\v c}}, M., {Allen}, S.~W., {Treu}, T., {Ebeling}, H., {Massey}, R.,
  {Morris}, R.~G., {von der Linden}, A., \& {Applegate}, D. 2008, \apj, 687,
  959

\bibitem[{{Cavaliere} \& {Fusco-Femiano}(1976)}]{cavaliere&ff76}
{Cavaliere}, A. \& {Fusco-Femiano}, R. 1976, \aap, 49, 137

\bibitem[{{Clowe} {et~al.}(2006){Clowe}, {Brada{\v c}}, {Gonzalez},
  {Markevitch}, {Randall}, {Jones}, \& {Zaritsky}}]{clowe.etal.06}
{Clowe}, D., {Brada{\v c}}, M., {Gonzalez}, A.~H., {Markevitch}, M., {Randall},
  S.~W., {Jones}, C., \& {Zaritsky}, D. 2006, \apjl, 648, L109

\bibitem[{{Cunha} \& {Evrard}(2009)}]{cunha&evrard}
{Cunha}, C.~E. \& {Evrard}, A.~E. 2009, ArXiv e-prints

\bibitem[{{Dolag} {et~al.}(2009){Dolag}, {Borgani}, {Murante}, \&
  {Springel}}]{dolag.etal.09}
{Dolag}, K., {Borgani}, S., {Murante}, G., \& {Springel}, V. 2009, \mnras, 399,
  497

\bibitem[{{Dolag} {et~al.}(2005){Dolag}, {Vazza}, {Brunetti}, \&
  {Tormen}}]{Dolag.etal.05a}
{Dolag}, K., {Vazza}, F., {Brunetti}, G., \& {Tormen}, G. 2005, \mnras, 994

\bibitem[{{Evrard}(1990)}]{evrard90}
{Evrard}, A.~E. 1990, \apj, 363, 349

\bibitem[{{Evrard} {et~al.}(2008){Evrard}, {Bialek}, {Busha}, {White}, {Habib},
  {Heitmann}, {Warren}, {Rasia}, {Tormen}, {Moscardini}, {Power}, {Jenkins},
  {Gao}, {Frenk}, {Springel}, {White}, \& {Diemand}}]{evrard.etal.08}
{Evrard}, A.~E., {Bialek}, J., {Busha}, M., {White}, M., {Habib}, S.,
  {Heitmann}, K., {Warren}, M., {Rasia}, E., {Tormen}, G., {Moscardini}, L.,
  {Power}, C., {Jenkins}, A.~R., {Gao}, L., {Frenk}, C.~S., {Springel}, V.,
  {White}, S.~D.~M., \& {Diemand}, J. 2008, \apj, 672, 122

\bibitem[{{Evrard} {et~al.}(1996){Evrard}, {Metzler}, \&
  {Navarro}}]{evrard.etal.96}
{Evrard}, A.~E., {Metzler}, C.~A., \& {Navarro}, J.~F. 1996, \apj, 469, 494

\bibitem[{{Fruscione} {et~al.}(2006){Fruscione}, {McDowell}, {Allen},
  {Brickhouse}, {Burke}, {Davis}, {Durham}, {Elvis}, {Galle}, {Harris},
  {Huenemoerder}, {Houck}, {Ishibashi}, {Karovska}, {Nicastro}, {Noble},
  {Nowak}, {Primini}, {Siemiginowska}, {Smith}, \& {Wise}}]{fruscione.etal.06}
{Fruscione}, A., {McDowell}, J.~C., {Allen}, G.~E., {Brickhouse}, N.~S.,
  {Burke}, D.~J., {Davis}, J.~E., {Durham}, N., {Elvis}, M., {Galle}, E.~C.,
  {Harris}, D.~E., {Huenemoerder}, D.~P., {Houck}, J.~C., {Ishibashi}, B.,
  {Karovska}, M., {Nicastro}, F., {Noble}, M.~S., {Nowak}, M.~A., {Primini},
  F.~A., {Siemiginowska}, A., {Smith}, R.~K., \& {Wise}, M. 2006, 6270

\bibitem[{{Gardini} {et~al.}(2004){Gardini}, {Rasia}, {Mazzotta}, {Tormen}, \&
  {Moscardini}}]{xmas}
{Gardini}, A., {Rasia}, E., {Mazzotta}, P., {Tormen}, G. aùnd~{De Grandi}, S.,
  \& {Moscardini}, L. 2004, \mnras, 351, 505

\bibitem[{{Haardt} \& {Madau}(1996)}]{haardt&madau}
{Haardt}, F. \& {Madau}, P. 1996, \apj, 461, 20

\bibitem[{{Hartley} {et~al.}(2008){Hartley}, {Gazzola}, {Pearce}, {Kay}, \&
  {Thomas}}]{hartley.etal.08}
{Hartley}, W.~G., {Gazzola}, L., {Pearce}, F.~R., {Kay}, S.~T., \& {Thomas},
  P.~A. 2008, \mnras, 386, 2015

\bibitem[{{Jeltema} {et~al.}(2008){Jeltema}, {Hallman}, {Burns}, \&
  {Motl}}]{jeltema.etal.08}
{Jeltema}, T.~E., {Hallman}, E.~J., {Burns}, J.~O., \& {Motl}, P.~M. 2008,
  \apj, 681, 167

\bibitem[{{Kravtsov} {et~al.}(2006){Kravtsov}, {Vikhlinin}, \&
  {Nagai}}]{kravtsov.etal.06}
{Kravtsov}, A.~V., {Vikhlinin}, A., \& {Nagai}, D. 2006, \apj, 650, 128

\bibitem[{{Lagan{\'a}} {et~al.}(2009){Lagan{\'a}}, {de Souza}, \&
  {Keller}}]{lagana.etal.09}
{Lagan{\'a}}, T.~F., {de Souza}, R.~S., \& {Keller}, G.~R. 2009, ArXiv e-prints

\bibitem[{{Lau} {et~al.}(2009){Lau}, {Kravtsov}, \& {Nagai}}]{lau.etal.09}
{Lau}, E.~T., {Kravtsov}, A.~V., \& {Nagai}, D. 2009, \apj, 705, 1129

\bibitem[{{Lima} \& {Hu}(2005)}]{lima&hu}
{Lima}, M. \& {Hu}, W. 2005, \prd, 72, 043006

\bibitem[{{Majumdar} \& {Mohr}(2004)}]{majumdar&mohr}
{Majumdar}, S. \& {Mohr}, J.~J. 2004, \apj, 613, 41

\bibitem[{{Mantz} {et~al.}(2009){Mantz}, {Allen}, {Rapetti}, \&
  {Ebeling}}]{mantz.etal.09}
{Mantz}, A., {Allen}, S.~W., {Rapetti}, D., \& {Ebeling}, H. 2009, astro-ph

\bibitem[{{Markevitch}(1998)}]{markevitch98}
{Markevitch}, M. 1998, \apj, 504, 27

\bibitem[{{Markevitch} {et~al.}(2002){Markevitch}, {Gonzalez}, {David},
  {Vikhlinin}, {Murray}, {Forman}, {Jones}, \& {Tucker}}]{markevitch.etal.02}
{Markevitch}, M., {Gonzalez}, A.~H., {David}, L., {Vikhlinin}, A., {Murray},
  S., {Forman}, W., {Jones}, C., \& {Tucker}, W. 2002, \apjl, 567, L27

\bibitem[{{Mazzotta} {et~al.}(2004){Mazzotta}, {Rasia}, {Moscardini}, \&
  {Tormen}}]{tsl}
{Mazzotta}, P., {Rasia}, E., {Moscardini}, L., \& {Tormen}, G. 2004, \mnras,
  354, 10

\bibitem[{{Meneghetti} {et~al.}(2009){Meneghetti}, {Rasia}, {Merten},
  {Bellagamba}, {Ettori}, {Mazzotta}, \& {Dolag}}]{meneghetti.etal.09}
{Meneghetti}, M., {Rasia}, E., {Merten}, J., {Bellagamba}, F., {Ettori}, S.,
  {Mazzotta}, P., \& {Dolag}, K. 2009, ArXiv e-prints

\bibitem[{{Nagai} {et~al.}(2007){Nagai}, {Vikhlinin}, \&
  {Kravtsov}}]{nagai.etal.07}
{Nagai}, D., {Vikhlinin}, A., \& {Kravtsov}, A.~V. 2007, \apj, 655, 98

\bibitem[{{O'Hara} {et~al.}(2006){O'Hara}, {Mohr}, {Bialek}, \&
  {Evrard}}]{ohara.etal.06}
{O'Hara}, T.~B., {Mohr}, J.~J., {Bialek}, J.~J., \& {Evrard}, A.~E. 2006, \apj,
  639, 64

\bibitem[{{Osmond} \& {Ponman}(2004)}]{osmond&ponman}
{Osmond}, J.~P.~F. \& {Ponman}, T.~J. 2004, \mnras, 350, 1511

\bibitem[{{Peebles}(1993)}]{peebles93}
{Peebles}, P.~J.~E. 1993, {Principles of physical cosmology} (Princeton Series
  in Physics, Princeton, NJ: Princeton University Press, |c1993)

\bibitem[{{Piffaretti} \& {Valdarnini}(2008)}]{piffa&valda}
{Piffaretti}, R. \& {Valdarnini}, R. 2008, \aap, 491, 71

\bibitem[{{Poole} {et~al.}(2007){Poole}, {Babul}, {McCarthy}, {Fardal},
  {Bildfell}, {Quinn}, \& {Mahdavi}}]{poole.etal.07}
{Poole}, G.~B., {Babul}, A., {McCarthy}, I.~G., {Fardal}, M.~A., {Bildfell},
  C.~J., {Quinn}, T., \& {Mahdavi}, A. 2007, \mnras, 380, 437

\bibitem[{{Pratt} {et~al.}(2010){Pratt}, {Arnaud}, {Piffaretti},
  {B{\"o}hringer}, {Ponman}, {Croston}, {Voit}, {Borgani}, \&
  {Bower}}]{pratt.etal.10}
{Pratt}, G.~W., {Arnaud}, M., {Piffaretti}, R., {B{\"o}hringer}, H., {Ponman},
  T.~J., {Croston}, J.~H., {Voit}, G.~M., {Borgani}, S., \& {Bower}, R.~G.
  2010, \aap, 511, A85+

\bibitem[{{Pratt} {et~al.}(2009){Pratt}, {Croston}, {Arnaud}, \&
  {B{\"o}hringer}}]{pratt.etal.09}
{Pratt}, G.~W., {Croston}, J.~H., {Arnaud}, M., \& {B{\"o}hringer}, H. 2009,
  \aap, 498, 361

\bibitem[{{Randall} {et~al.}(2002){Randall}, {Sarazin}, \&
  {Ricker}}]{randall.etal.02}
{Randall}, S.~W., {Sarazin}, C.~L., \& {Ricker}, P.~M. 2002, \apj, 577, 579

\bibitem[{{Rasia} {et~al.}(2006){Rasia}, {Ettori}, {Moscardini}, {Mazzotta},
  {Borgani}, {Dolag}, {Tormen}, {Cheng}, \& {Diaferio}}]{rasia.etal.06}
{Rasia}, E., {Ettori}, S., {Moscardini}, L., {Mazzotta}, P., {Borgani}, S.,
  {Dolag}, K., {Tormen}, G., {Cheng}, L.~M., \& {Diaferio}, A. 2006, \mnras,
  369, 2013

\bibitem[{{Rasia} {et~al.}(2008){Rasia}, {Mazzotta}, {Bourdin}, {Borgani},
  {Tornatore}, {Ettori}, {Dolag}, \& {Moscardini}}]{rasia.etal.08}
{Rasia}, E., {Mazzotta}, P., {Bourdin}, H., {Borgani}, S., {Tornatore}, L.,
  {Ettori}, S., {Dolag}, K., \& {Moscardini}, L. 2008, \apj, 674, 728

\bibitem[{{Rasia} {et~al.}(2004){Rasia}, {Tormen}, \& {Moscardini}}]{rtm}
{Rasia}, E., {Tormen}, G., \& {Moscardini}, L. 2004, \mnras, 351, 237

\bibitem[{{Ricker} \& {Sarazin}(2001)}]{ricker&sarazin}
{Ricker}, P.~M. \& {Sarazin}, C.~L. 2001, \apj, 561, 621

\bibitem[{{Ritchie} \& {Thomas}(2002)}]{ritchie&thomas}
{Ritchie}, B.~W. \& {Thomas}, P.~A. 2002, \mnras, 329, 675

\bibitem[{{Rowley} {et~al.}(2004){Rowley}, {Thomas}, \& {Kay}}]{rowley.etal.04}
{Rowley}, D.~R., {Thomas}, P.~A., \& {Kay}, S.~T. 2004, \mnras, 352, 508

\bibitem[{{Short} {et~al.}(2010){Short}, {Thomas}, {Young}, {Pearce},
  {Jenkins}, \& {Muanwong}}]{short.etal.10}
{Short}, C.~J., {Thomas}, P.~A., {Young}, O.~E., {Pearce}, F.~R., {Jenkins},
  A., \& {Muanwong}, O. 2010, ArXiv e-prints

\bibitem[{{Springel}(2005)}]{gadget2}
{Springel}, V. 2005, \mnras, 364, 1105

\bibitem[{{Springel} \& {Hernquist}(2003)}]{springel&hernquist03}
{Springel}, V. \& {Hernquist}, L. 2003, \mnras, 339, 289

\bibitem[{{Stanek} {et~al.}(2009){Stanek}, {Rasia}, {Evrard}, {Pearce}, \&
  {Gazzola}}]{stanek.etal.09}
{Stanek}, R., {Rasia}, E., {Evrard}, A.~E., {Pearce}, F., \& {Gazzola}, L.
  2009, ArXiv e-prints

\bibitem[{{Tormen} {et~al.}(1997){Tormen}, {Bouchet}, \&
  {White}}]{Tormen.etal.97}
{Tormen}, G., {Bouchet}, F.~R., \& {White}, S.~D.~M. 1997, \mnras, 286, 865

\bibitem[{{Tornatore} {et~al.}(2007){Tornatore}, {Borgani}, {Dolag}, \&
  {Matteucci}}]{tornatore.etal.07}
{Tornatore}, L., {Borgani}, S., {Dolag}, K., \& {Matteucci}, F. 2007, \mnras

\bibitem[{{Torri} {et~al.}(2004){Torri}, {Meneghetti}, {Bartelmann},
  {Moscardini}, {Rasia}, \& {Tormen}}]{torri.etal.04}
{Torri}, E., {Meneghetti}, M., {Bartelmann}, M., {Moscardini}, L., {Rasia}, E.,
  \& {Tormen}, G. 2004, \mnras, 349, 476

\bibitem[{{Vikhlinin} {et~al.}(2006){Vikhlinin}, {Kravtsov}, {Forman}, {Jones},
  {Markevitch}, {Murray}, \& {Van Speybroeck}}]{vikh.etal.06}
{Vikhlinin}, A., {Kravtsov}, A., {Forman}, W., {Jones}, C., {Markevitch}, M.,
  {Murray}, S.~S., \& {Van Speybroeck}, L. 2006, \apj, 640, 691

\bibitem[{{Vikhlinin} {et~al.}(2009){Vikhlinin}, {Kravtsov}, {Burenin},
  {Ebeling}, {Forman}, {Hornstrup}, {Jones}, {Murray}, {Nagai}, {Quintana}, \&
  {Voevodkin}}]{vikh.etal.09}
{Vikhlinin}, A., {Kravtsov}, A.~V., {Burenin}, R.~A., {Ebeling}, H., {Forman},
  W.~R., {Hornstrup}, A., {Jones}, C., {Murray}, S.~S., {Nagai}, D.,
  {Quintana}, H., \& {Voevodkin}, A. 2009, \apj, 692, 1060

\bibitem[{{Vikhlinin} {et~al.}(1998){Vikhlinin}, {McNamara}, {Forman}, {Jones},
  {Quintana}, \& {Hornstrup}}]{vikh.etal.98}
{Vikhlinin}, A., {McNamara}, B.~R., {Forman}, W., {Jones}, C., {Quintana}, H.,
  \& {Hornstrup}, A. 1998, \apj, 502, 558

\bibitem[{{Wik} {et~al.}(2008){Wik}, {Sarazin}, {Ricker}, \&
  {Randall}}]{wik.etal.08}
{Wik}, D.~R., {Sarazin}, C.~L., {Ricker}, P.~M., \& {Randall}, S.~W. 2008,
  \apj, 680, 17

\bibitem[{{Yang} {et~al.}(2009){Yang}, {Ricker}, \& {Sutter}}]{yang.etal.09}
{Yang}, H., {Ricker}, P.~M., \& {Sutter}, P.~M. 2009, \apj, 699, 315

\bibitem[{{Yoshida} {et~al.}(2001){Yoshida}, {Sheth}, \&
  {Diaferio}}]{yoshida.etal.01}
{Yoshida}, N., {Sheth}, R.~K., \& {Diaferio}, A. 2001, \mnras, 328, 669

\end{thebibliography}

\end{document}